\documentclass[epj]{svjour}
\usepackage{graphics}
\usepackage{amsmath}    
\usepackage{amssymb}    
\usepackage{bm}
\usepackage{graphicx, stackengine}   
\usepackage{verbatim}   
\usepackage{color}      
\usepackage{subfigure}  
\usepackage{hyperref} 
\usepackage{appendix} 
\usepackage{cuted}
\usepackage{capt-of}
\newcommand{\mb}{\mathbf}

\begin{document}

\title{Mobilities of a drop and an encapsulated squirmer}
\author{R. Kree \and  A.Zippelius}
\institute{Georg-August-Universit\"at G\"ottingen, 
Institut f\"ur Theoretische Physik, Friedrich-Hund-Platz 1, 37077 G\"ottingen, Germany
\email{kree@theorie.physik.uni-goettingen.de}}
\date{Received: \today / Revised version: date}

\abstract{
  We have analyzed the dynamics of a spherical, uni-axial squirmer which is
  located inside a spherical liquid drop at general position $\bm{r}_s$.  The squirmer is subject to an
  external force and torque in addition to the slip velocity on its
  surface.  We have derived exact analytical expressions for the
  linear and rotational velocity of the squirmer as well as the linear
  velocity of the drop for general, non-axisymmetric configurations. The mobilities of both, squirmer and
  drop, are in general anisotropic, depending on the orientation of
  $\bm{r}_s$, relative to squirmer axis, external force or torque.  We
  discuss their dependence on the size of the squirmer, its distance
  from the center of the drop and the viscosities.  Our
  results provide a first step towards a discussion of the trajectories
  of the composite system of drop and enclosed squirmer.   
}

\maketitle


\maketitle

\section{Introduction}

Controlled locomotion on micro- or nanometer scales is of great
interest for both, cell biology and
micro-robotics~\cite{Lauga2009,Ramaswamy2010,Marchetti2013,Bechinger2016a,Juelicher2018,Wang2021}. In the former case, one
aims to understand the swimming motion of microorganisms and cell
motility. In the latter case, the goals are control and design of
micro-robots optimized for a variety of biomedical applications.  Our
focus here is on a composite system, consisting of an active device,
encapsulated in a liquid drop.

Such composite systems have been studied experimentally in several
different setups, using liquid droplets containing concentrated
aqueous solution of
bacteria~\cite{Wioland2013,Soto2020,Lavrentovich2021,Clement2019} in
order to understand pattern formation and swimming in a confined
geometry. One example are suspensions of {\it Bacillus subtilis} which
form stable vortex patterns inside a liquid
drop~\cite{Wioland2013}. In another setup {\it Escherichia coli} in a water oil
emulsion was shown to be able to propel the
droplet~\cite{Soto2020}. Similar propulsion has been observed for bacteria in a liquid droplet, when put into an ordered liquid crystalline state
with defects~\cite{Lavrentovich2021}. Yet another example are
magnetotactic bacteria which were shown to selfassemble into a rotary
motor~\cite{Clement2019}.
In the context of micro-robotics, synthetic microswimmers, such as
artificial bacterial flagella~\cite{deMello2016} or photocatalytic particles~\cite{Dietrich2020} are able to
propel liquid droplets, which is of interest in many biomedical
applications, such as targeted drug delivery.  The big
 advantage of
self-propulsion is that energy can be supplied by the
 surroundings;
the main disadvantage is lack of control. Therefore
 a combination of
both, self-propulsion and actuation by external fields, is a promising
candidate to achieve
 optimal control of an otherwise self-propelled
composite device.

Most theoretical studies of composite systems have focussed on simple
internal active devices. The simplest ones are point
forces~\cite{Rueckert2021,Daddi2020}, which can be combined to model
pullers and pushers. Alternatively the active device has been taken as
a squirmer~\cite{Reigh2017} whose slip velocity generates a flow
inside the droplet and thereby can propel
it~\cite{Lighthill1952,Blake1971}. Marangoni flow on the droplet`s
surface provides another driving mechanism, leading to stable comoving
states~\cite{Shaik2018}. In yet another approach, the device is a
passive particle encapsulated in the droplet, experiencing external
forcing or shear flow~\cite{Thampi2019}.  In all these studies analytical solutions were
given for the concentric configuration only.  The more complex system
with many squirmers inside a droplet was studied numerically in
ref.~\cite{Huang2020}; propulsion of the droplet was observed only, if
the encapsulated squirmers moved coherently.

In a previous paper~\cite{Kree2021a}, henceforth denoted by I, we
presented an analytical solution for a squirmer, encapsulated in a
drop and displaced from the center of the drop by ${\bm a}$. We
only discussed the axisymmetric case, such that both, the symmetry
axis of the squirmer and an applied external force, are parallel to
the displacement ${\bm a}$. We identified stable, co-moving states of
squirmer and drop which can be achieved by an appropriate
adjustment of the external force such that squirmer and drop move
with the same velocity. These states allow for a controlled
manipulation of the viscous drop by external forcing. 

Here we extend our analysis and calculate the mobilities of both the squirmer and the drop for general orientations of the
displacement $\bm a$ with respect to the symmetry axis of the squirmer
and/or the applied external force. For the non-collinear arrangement, the squirmer is subject to a torque with respect to the center of the drop and hence rotates in addition to its linear velocity. We also include an applied external torque, which might be
generated by an external magnetic field, provided the active particle
is magnetized. In fact propulsion of helical structures by rotating
magnetic fields has been discussed in
detail~\cite{Tottori2012,Morozov2014,Servant2015}, and biohybrid helical spermbots are interesting candidates for biomedical applications ~\cite{Medina2017}.
Electric fields
could also provide a torque, if the active particle has a permanent
dipole moment.

The linearity of Stokes equation allows us to decompose the analytical calculations into subproblems. We first solve  
(i) an autonomous swimmer, (ii) a passive particle, which is driven by   (iia) an external force  or (iib) an external torque.   The case of an
encapsulated squirmer, subject to an external force and torque is
obtained by superpositions of (i), (iia) and (iib). The analytical solution is constructed in a special geometry, for which the displacement of the squirmer is perpendicular to the squirmer axis or the direction of external force. Then we superimpose this solution with that of reference I and use frame independence to obtain our results for general displacements and orientations . 

The paper is organized as follows: The model is
defined in sec.\ref{model}; the analytical method and the solution is presented in
sec.\ref{analytics}. The results of the analytical calculation are the
mobilities of the squirmer and the drop as functions of the sizes of particle and drop, the displacement vector and the viscosities.
 They are presented in sec.\ref{results1} and discussed in sec.\ref{discuss}. 

\section{Model}
\label{model}
\begin{figure}
\includegraphics[width=0.5\textwidth]{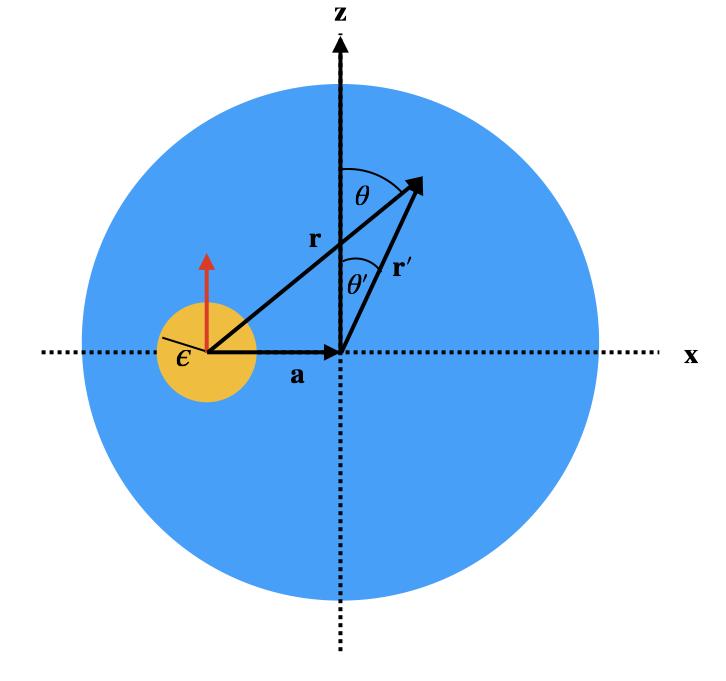}
\caption{\label{device} Geometry used in sect. \ref{model} and \ref{analytics}. Squirmer (yellow) of radius $\epsilon$, encapsulated in a
  viscous drop (blue) and displaced from the center by $\bm a$. The
  direction of the displacement is chosen perpendicular to the
  symmetry axis of the squirmer, $\bm n$, shown as a red arrow and chosen to point along ${\bm e}_z$. If external forces $\bm{F}_{ext}$ are present, they also point in z-direction and external torques $\bm{D}_{ext}$ point in y-direction.}
\end{figure}

We study the propulsion of a viscous drop, which is driven by an
active device in its interior, as depicted in Fig.\ref{device}. The
active device is either a squirmer with a tangential slip velocity on
its surface (1) or a passive particle, subject to an external force
${\bm F}^{ext}$ and/or torque ${\bm D}^{ext}$ (2), or a combination of both. The active device is
modeled as a solid particle of radius $\epsilon$, positioned at
${\bm r}_s=-\bm{a}$, measured from the center of the drop.
Following the above strategy, we consider {\it perpendicular}
alignment of $\bm{a}$ and squirmer axis ${\bm n}$ for problem (1) and
similarly {\it perpendicular} alignment of $\bm{a}$ and
${\bm F}_{ext}$ and $\bm{D}_{ext}$ for problem (2).  We first choose
special coordinates with $\bm{a}=a\bm{e}_x$, $\bm{n}=\bm{e}_z$ and ${\bm F}_{ext}=F_{ext}{\bm e}_z$ and $\bm{D}_{ext}=D_{ext}\bm{e}_y$. In all of this and the next section, we will stick to this assignment and postpone a discussion of general relative orientations to sec.~\ref{results1}. We introduce two frames of reference: one with its origin in the center of the particle (P) and one with its origin in the center of the drop (D). A point has position vector $\bm r$ in the first frame and position vector ${\bm r}^{\prime}={\bm r}-{\bm a}$ in the second (see Fig\ref{device}).

The drop is assumed to be spherical and consists of an
incompressible Newtonian fluid with viscosity $\eta^-$. It is immersed
in an ambient Newtonian fluid of viscosity $\eta^+$ which is at rest
in the laboratory frame (LF). The two fluids are assumed to be
completely immiscible, and the drop is neutrally buoyant. We
choose units of mass, length and time such that the density
$\rho_0=1$, the drop radius $R=1$ and the viscosity of the exterior
fluid $\eta^+=1$. We do, however, keep the notation $\eta^+$, because some results, e.g. the mobility of the drop in the exterior fluid, are more intuitive with the explicit notation.
 The slip velocity is expanded in spherical harmonics $\chi_{lm}(\Omega)=P_l^m(\cos{\theta})e^{im\phi}$ ($P_l^m$ denoting the associated Legendre polynomials.
) in the coordinate system (P) of the squirmer, i.e. the angle $\Omega=(\theta, \phi)$ has its vertex at the center of the squirmer. 
 For the purposes of this work, we only have to consider $\ell=1$ components, and we choose the polar axis $\theta=0$  so that
the slip velocity on the surface of the squirmer is given by
\begin{equation}
\label{eq:slip}
  \bm{v}_{slip}(\Omega)= -h\sin\theta \bm{e}_\theta=h\nabla_s\chi_{10},
   \end{equation}
where $\nabla_s$ denotes the surface gradient.
 
 The squirmer generates a flow field inside ($\bm{v}^-$) and outside ($\bm{v}^+$) of the drop.
For small Reynolds number the flow fields can be calculated from Stokes's equation 
\begin{equation}
\label{eq:stokes}
\nabla\cdot\bm{\sigma}^{\pm}=\eta^{\pm}\nabla^2\bm{v}^{\pm}-\nabla p^{\pm}=0 , 
\end{equation}
and the incompressibility condition $\nabla\cdot\bm{v}^{\pm}=0$.
The viscous stress tensor $\bm{\sigma}^{\pm}$ is given by its
cartesian components
$\sigma_{ij}^{\pm}=-p^{\pm}\delta_{ij}+\eta^{\pm}(\partial_iv_j^{\pm}+\partial_jv_i^{\pm})$, with
the pressure $p$ determined from incompressibility.

Stokes equation has to be supplemented by boundary conditions on the
surface of the active particle and on the surface of the
drop. Given the displacement of the active particle away form the
center of the drop, we expect linear as well as rotational motion
of the particle (squirmer or dragged passive particle).  Hence
the flow field on the surface of the particle in frame (P) 
takes the general form
\begin{equation}
  \label{eq:bc1}
  \bm{v}^-(\bm{r})=\bm{v}_{slip}(\bm{r})+\bm{U}+\bm{\omega}\times \bm{r}
  \quad \text{for} \quad r=\epsilon,
\end{equation}
where $\bm{U}$ denotes the linear and $\omega$ the rotational velocity of the particle. Continuity of the flow field is assumed for points  on the
surface  of the drop in frame (D) with position vector ${\bm r}^{\prime}$
\begin{equation}
  \label{eq:bc2}
\bm{v}^-(\bm{r^{\prime}})=\bm{v}^+(\bm{r^{\prime}}) \quad \text{for}\quad r'=1.
\end{equation}
The tangential stress is continuous, whereas the normal
stress jumps due to a homogeneous surface tension $\gamma_0$, so that
\begin{equation}
  \label{eq:bc3}
\bm{e}_{r^{\prime}}\cdot(\bm{\sigma}_+-\bm{\sigma}_-)=2\gamma_0\bm{e}_{r^{\prime}}
\quad \text{for} \quad r'=1,
\end{equation}
with $\bm{e}_{r'}=\bm{r'}/r'$.
Once the flow fields $\bm{v}^{\pm}$ have been determined, the linear velocity of the drop can be computed as an integral over the drop's surface from
\begin{equation}
  \label{centerofmass}
  \bm{v}_{CM}=\frac{3}{4\pi}\int_{r'=1} d^2x \;(\bm{e}_{r^{\prime}}\cdot \bm{v}^{\pm})\,\bm{e}_{r^{\prime}}.
\end{equation}


\section{ Analytical Solution}
\label{analytics}

Our strategy for the analytical solution is analogous to the one previously used in I. We briefly recall it for consistency. In a first step, the internal flow, $\bm{v}^-$, is expanded in a
complete set of solutions of the Stokes equations in frame (P),
and is matched to the slip velocity on the squirmer's surface. This
solution is similar to the flow field of a squirmer in unbounded
space, but contains --- in addition to fields which are regular
at infinity--- also those which are regular at the squirmer's center and would be forbidden in unbounded space.

The boundary conditions on the drop's surface are easily formulated
in the frame (D). Therefore we seek  to expand the flow field $\bm{v}^-(\bm{r})$$ =\bm{v}^-(\bm{r'} + \bm{a})$ around the drop's center in the same set of solutions as used in (P).  In contrast to reference I,
we consider displacements of the squirmer 
which are perpendicular to either the symmetry axis of the squirmer or the applied external force or torque.


\subsection{Solutions of Stokes equation}
In spherical geometries it is advantageous to construct solutions of the Stokes equations from vector spherical harmonics. We use the following set of functions, which diagonalize the surface Laplacian:  
\begin{eqnarray}
  \bm{\Psi}^1_{lm}&=&\nabla_s\chi_{lm}+l\chi_{lm} \bm{e}_r\\
  \bm{\Psi}^3_{lm}&=&\nabla_s\chi_{lm}-(l+1)\chi_{lm} \bm{e}_r\\
\bm{\Psi}^2_{lm}&=& \bm{e}_r\times\nabla_s\chi_{lm}.
\end{eqnarray}


Here, $l$ and $m$ are integers with $0\leq l$ and $-l\leq m \leq l$ and $\nabla_s$ denotes the surface gradient.  
Solutions of the Stokes equations can be
classified according to whether they are regular at the origin (inner solutions) or
regular at infinity (outer solutions). 
We use a complete set of inner solutions given by
\begin{eqnarray*}
  \label{ukleiner}
  \bm{u}_{lm}^{1<}&=&\frac{r^{l-1}}{(l+m)!}  \bm{\Psi}^1_{lm}\\
\bm{u}_{lm}^{2<}&=&\frac{r^{l}}{(l+m)!(l+1)}  \bm{\Psi}^2_{lm}\\
\bm{u}_{lm}^{3<}&=&\frac{r^{l+1}}{(l+m)!(2l+1)} (\bm{\Psi}^1_{lm}+\frac{2l}{(l+1)(2l+3)} \bm{\Psi}^3_{lm}) , 
  \end{eqnarray*} 
whereas outer solutions will be used in the form
\begin{eqnarray*}
  \label{ugroesser}
  \mb{u}_{lm}^{1>}&=&\frac{(l-m)!}{r^{l+2}}  \bm{\Psi}^3_{lm}\\
\bm{u}_{lm}^{2>}&=&-\frac{(l-m)!}{lr^{l+1}}  \bm{\Psi}^2_{lm}\\
  \bm{u}_{lm}^{3>}&=&\frac{(l-m)!}{r^l(2l+1)} (-\bm{\Psi}^3_{lm}
                      +\frac{2(l+1)}{l(2l-1)} \bm{\Psi}^1_{lm}) . 
\end{eqnarray*}

The slip velocity  Eq. (\ref{eq:slip}) takes on the form  $\bm{v}_{slip}=
h \nabla_s\chi_{10}= (h/3) (\bm{\psi}^3_{10} + 2\bm{\psi}^1_{10})$ in vector spherical harmonics. 
For displacements $\bm{a}=a\bm{e}_x$,  we expect to find a solution with $\bm{U}=U\bm{e}_z$
and $\bm{\omega}=\omega\bm{e}_y$. These velocities are expanded in vector spherical harmonics as follows:
\begin{align}
  \bm{U}&=U \bm{\Psi}^1_{10}\\
  \bm{\omega}\times\bm{r} & =-\frac{i}{2}\omega r (\bm{\Psi}^2_{11}+
  2\bm{\Psi}^2_{1-1})=r\omega \Im \bm{\Psi}^2_{11},
\end{align}
where $\Im$ denotes the imaginary part.
To construct a solution of the boundary value problem, we start from an Ansatz with three vector spherical harmonics: $\bm{\Psi}^1:=\bm{\Psi}^1_{10}$, $\bm{\Psi}^3:=\bm{\Psi}^3_{10}$ and $\bm{\Psi}^2:=\Im \bm{\Psi}^2_{11}=(\bm{\Psi}^2_{11}+2\bm{\Psi}^2_{11-})/(2i)$.
The inner and outer solutions take on the form:
\begin{eqnarray}
  \label{u1<}
  \bm{u}^{1<}&:=&\bm{u}^{1<}_{10}=\bm{\Psi}^1\\
   \label{u2<}
  \bm{u}^{2<}&:=& (\bm{u}^{2<}_{11}+2\bm{u}^{2<}_{1-1})/(2i)=\frac{r}{4}\bm{\Psi}^2\\
   \label{u3<}
  \bm{u}^{3<}&:=&\bm{u}^{3<}_{10}=\frac{r^2}{3} (\bm{\Psi}^1+\frac{1}{5} \bm{\Psi}^3)\\
   \label{u1>}
  \bm{u}^{1>}&:=&\bm{u}^{1>}_{10}=\frac{1}{r^3}  \bm{\Psi}^3\\
   \label{u2>}
  \bm{u}^{2>}&:=& (\bm{u}^{2>}_{11}+2\bm{u}^{2>}_{1-1})/(2i)=-\frac{1}{r^{2}} \bm{\Psi}^2\\
   \label{u3>}
  \bm{u}^{3>}&:=&\bm{u}^{3>}_{10}=\frac{1}{3r} (-\bm{\Psi}^3
                  +4\bm{\Psi}^1) . 
\end{eqnarray}
The only solutions, which are accompanied by pressure are $\bm{u}^{3>}$ and $\bm{u}^{3<}$, and the corresponding pressures are explicitly given by
\begin{equation}
\label{eq:pressure}
  p^<= 2\eta^-r\cos{\theta} \qquad \text{and} \qquad p^>= 2\eta^-\cos{\theta}/r^2,
\end{equation}
 apart from a constant reference pressure.
 
The general solution of Stokes equation in frame (P)
in the interior of the drop is given by a superposition of both, the inner and outer solutions:
\begin{equation}
  \label{eq:v-}
  \bm{v}^-=a_1\bm{u}^{1<}+a_2 \bm{u}^{2<}+a_3\bm{u}^{3<}+b_1\bm{u}^{1>}
  +b_2 \bm{u}^{2>}+b_3\bm{u}^{3>}.
\end{equation}
The flow field in frame (D) outside of the drop  is given by 
\begin{equation}
\label{eq:exteriorvelocity}
  \bm{v}^+=c_1\bm{u}^{1>}++c_2\bm{u}^{2>}+ c_3\bm{u}^{3>}.
\end{equation}
This Ansatz involves nine parameters, which have to be
determined by the boundary conditions.

\subsection{Drop velocity}
To express the drop velocity $\bm{v}_{cm}$ by the parameters of the solution, we insert Eq. (\ref{eq:exteriorvelocity}) into 
Eq.~(\ref{centerofmass}).  With $\bm{e}_r\cdot\bm{\Psi^1}=\cos\theta$ and $\bm{e}_r\cdot\bm{\Psi}^3=-2\cos\theta$ we find
\begin{equation}
\label{eq:vcm}
\bm{v}_{CM}=v_{CM}\bm{e}_z=2(c_3-c_1)\bm{e}_z.
\end{equation}

\subsection{Boundary condition on the surface of the squirmer}
\label{bcsquirmer}
Given the interior flow field in the form of an expansion around the
squirmer`s center~(\ref{eq:v-}) in frame (P), we can easily fulfill the boundary condition on the squirmer's surface.  Plugging our Ansatz into
Eq.(\ref{eq:bc1}), we obtain three equations:
\begin{align}
  a_1+\frac{\varepsilon^2}{3}a_3+\frac{4}{3\varepsilon}b_3=&U+\frac{2}{3}h \label{eq:linbc1}\\
  \frac{\varepsilon^2}{15}a_3+\frac{1}{\varepsilon^3}b_1-\frac{1}{3\varepsilon}b_3=&\frac{1}{3}h.\label{eq:linbc2}\\
  \frac{\varepsilon}{4}a_2-\frac{b_2}{\varepsilon^2}=&\varepsilon\omega, \label{eq:linbcomega}
\end{align}
relating the coefficients of the interior flow field to the activity
of the squirmer.
Further six equations are provided by the boundary conditions on the drop's surface Eqs.(\ref{eq:bc2},\ref{eq:bc3}). However, before we can use them, we have to shift the internal flow $\bm{v}^-$, given in frame (P) to its representation in frame (D).\\

\subsection{Translations}
To express the vector field $\bm{v}^-$, given in terms of the solutions Eqs.(\ref{u1<}-\ref{u3>}) in frame (P), on the surface of the drop in frame (D), we derive a generalization of the corresponding transformations
for the solid scalar spherical harmonics, which can be found in in
ref.\cite{VanGelderen1998}. These translations are easily worked out by hand for the $\ell=1$ components of the flow:
\begin{eqnarray*}
  \label{l=1m=0}
  \bm{u}^{1<}(\bm{r'}+a\bm{e}_x)&=&  \bm{u}^{1<}(\bm{r'})\\
  \bm{u}^{2<}(\bm{r'}+a\bm{e}_x)&=&\bm{u}^{2<}(\bm{r'})
                                      - \frac{a}{4}\bm{u}^{1<}(\bm{r'})+\cdots\\
  \bm{u}^{3<}(\bm{r'}+a\bm{e}_x)&=&\bm{u}^{3<}(\bm{r'})+
                            \frac{2a^2}{5}\bm{u}^{1<}(\bm{r'})-2a \bm{u}^{2<}(\bm{r'})+\cdots\\
  \bm{u}^{1>}(\bm{r'}+a\bm{e}_x)&=&\bm{u}^{1>}(\bm{r'})+\cdots\\
 \bm{u}^{2>}(\bm{r'}+a\bm{e}_x)&=&\bm{u}^{2>}(\bm{r'})
                                      - \frac{a}{2}\bm{u}^{1>}(\bm{r'})+\cdots\\ 
  \bm{u}^{3>}(\bm{r'}+a\bm{e}_x)&=&\bm{u}^{3>}(\bm{r'})+
                            \frac{2a^2}{5}\bm{u}^{1>}(\bm{r'})-a \bm{u}^{2>}(\bm{r'})+\cdots
\end{eqnarray*}
The ellipsis denote non-vanishing terms with $l\geq 2$. These terms do not contribute to the velocities of squirmer and drop, but will in general lead to deformations of the drop's spherical shape. In the present work we do not study this part of the flow. 

The pressure  is expanded in terms of scalar spherical harmonics and it turns out that it remains unchanged,  $p(\bm{r'}+a\bm{e}_x) = p(\bm{r'})$.

\subsection{Boundary conditions on the surface of the drop}
\label{bcdrop}
Given the translated velocity fields, we can evaluate the internal flow
$\bm{v}^-(\bm{e}_{r^{\prime}}+a\bm{e}_x)$ at the boundary of the drop ($r'=1$), as needed for the second boundary condition Eq.(\ref{eq:bc2}). Continuity of the velocity across the drop's surface implies
\begin{align}
  a_1+\frac{a_3}{3}(1+\frac{6}{5} a^2)+\frac{4}{3}(b_3-c_3)-\frac{a}{4}a_2&=0 \label{eq:linbc3}\\
  b_1-\frac{b_3}{3}(1-\frac{6}{5} a^2)+\frac{a_3}{15}-\frac{a}{2}b_2-c_1+\frac{c_3}{3}&=0\\
-\frac{a}{2}a_3+\frac{a_2}{4}+ab_3-b_2+c_2&=0.
  \label{eq:linbc4}
\end{align}

To fulfill the balance of forces on the drop's surface
Eq.(\ref{eq:bc3}), we need to compute the  tractions $\bm{t}=-p\bm{e}_r + \bm{t}_{vis}$. The viscous part is obtained for any Stokes flow $\bm{u}^\pm$, using the identity

\begin{equation}
  \bm{t}_{vis}^{\pm}=\eta^{\pm}\big(2 \frac{\partial}{\partial r}+{\bm e}_r\times
  \nabla\times\big){\bm u}^{\pm}.
\end{equation}
Together with the pressure contribution, we find for the tractions in the interior of the fluid
  \begin{eqnarray}
    \bm{t}^{1<}&=&0\\
    \bm{t}^{2<}&=&0\\
    \bm{t}^{3<}&=&\frac{3\eta^-r}{5}\bm{\Psi}^3\\
    \bm{t}^{1>}&=& -\frac{6\eta^-}{r^4}\bm{\Psi}^3\\
    \bm{t}^{2>}&=& \frac{3\eta^-}{r^3}\bm{\Psi}^2\\
    \bm{t}^{3>}&=&\frac{2\eta^-}{r^2}(\bm{\Psi}^3-\bm{\Psi}^1),                  
  \end{eqnarray}
represented in the frame (P). Since the tractions are linear functions of the velocities, the transformation to the frame (D) are easily obtained from the transformation of the velocities:

\begin{eqnarray}
       \bm{t}^{2>}&=& =\frac{3\eta^-}{r^3}\bm{\Psi}^2+\frac{3a\eta^-}{r^4}\bm{\Psi}^3\\
    \bm{t}^{3>}&=&\frac{2\eta^-}{r^2}(\bm{\Psi}^3-\bm{\Psi}^1)-\frac{12a^2\eta^-}{5r^4}\bm{\Psi}^3  -\frac{3a\eta^-}{r^3 }\bm{\Psi}^2
 \end{eqnarray}
All other tractions turn out to be unaffected by the translation. 
  

  The above
  tractions are plugged into the third boundary condition
  Eq.~(\ref{eq:bc3}), implying 3 more linear equations for the yet unknown coefficients
\begin{eqnarray}
  b_3&=&\frac{c_3}{\lambda}\label{bcIII.1}\\
  \frac{3}{5}a_3-6b_1+3ab_2+2b_3(1-\frac{6}{5}a^2)&=&\frac{2(c_3-3c_1)}{\lambda}
\label{bcIII.2}  \\
b_2-ab_3&=&\frac{c_2}{\lambda\label{bcIII.3}},
\end{eqnarray}
where $\lambda=\eta^-/\eta^+$ denotes the viscosity contrast.
  

\subsection{Force and torque balance}

The boundary conditions on the surface of the squirmer and the drop
provide nine linear equations. Force and torque balance yield two more
equations, so that all
unknowns, the nine coefficients of the general solution and $U$ and
$\omega$, are uniquely determined.

An external force acting on the particle has to be balanced by
the total viscous force: $\bm{F}_{visc}+\bm{F}_{ext}=\bm{0}$.
The total viscous force
  $\bm{F}_{visc}$ can be expressed as an integral of the tractions
  over the surface of a large sphere of radius $ R\gg 1$)
\begin{equation}
\bm{F}_{visc}=\lim_{R\to \infty} \int_{R} d^2 x \;\bm{t}^+.
\end{equation}
The only flow term contributing to this expression is
$c_3\bm{u}^{3>}$ which is $\sim 1/r$, so that
\begin{equation}\label{forcebalance}
\bm{F}_{visc}= -8\pi\eta^+ c_3\bm{e}_z=-\bm{F}_{ext}.
\end{equation}
Hence force balance determines the coefficient 
$c_3$. 

In the balance of torque $\bm{D}_{visc} + \bm{D}=\bm{0}$ the viscous part is determined  from 
\begin{equation}
\bm{D}_{visc}=\lim_{R\to\infty} \int_R d^2 x \;\bm{r}\times\bm{t}^+.
\end{equation}
Note that this torque is calculated in frame (D). The only flow term contributing to this expression is
$c_2\bm{u}^{2>}$ which falls off as $\sim 1/r^2$, so that in our geometry $\bm{D}_{visc}=D_{visc}\bm{e}_y$ with
\begin{equation}\label{torquebalance}
  D_{visc}=8\pi\eta^+c_2\bm{e}_y.
\end{equation}
The exerted torque in this frame is given by $\bm{D}=\bm{D}_F + \bm{D}_{ext}$. The first term $\bm{D}_F= -\bm{a}\times \bm{F}_{ext} $ arises from any moment-free force distribution with total force $\bm{F}_{ext}$. 
In our special geometry, the torque balance becomes 
\begin{equation}
8\pi\eta^+c_2 + aF_{ext} + D_{ext}=0,	
\end{equation} 
which fixes the parameter $c_2$.

\section{Mobilities of squirmer and drop}\label{results1}

The analytical solution of the linear system of Eqs. (\ref{eq:linbc1}- \ref{eq:linbcomega}), (\ref{eq:linbc3}-\ref{eq:linbc4}) and (\ref{bcIII.1}-\ref{bcIII.3}) is discussed here for three different situations:
\begin{itemize}
\item{an autonomous squirmer without applied external force or torque}
\item{a passive particle (no slip velocity) dragged by an external force}
  \item{a passive particle (no slip velocity) subject to an external torque.}
\end{itemize}
We extract the analytical expressions for the mobilities, relating
$\bm{F}_{ext}, \bm{D}_{ext}$ and the squirmer activity $h$ to the velocities
$\bm{U},\bm{\omega}$ of the particle and $\bm{v}_{cm}$ of the drop.
Combining these results with reference I and using frame independence,
we then obtain the mobility tensors of the particle and the drop for
each of the three cases. The complete analytical expressions for all
the mobility tensors, including those from reference I can be worked out by hand (and have been checked
by symbolic computing using \textit{SymPy} \cite{sympy}). They are
summarized in Appendix \ref{app:mobilities}. More general situations,
representing a squirmer subject to both, external force and torque,
which drives its enclosing drop, can be obtained by linear superposition. 
   
\subsection{Encapsulated squirmer}

The activity of the squirmer is conveniently characterized by its
velocity $U_0=-2h/3$ in an unbounded fluid.
For the autonomous swimmer, force and torque balance imply $b_3=c_3=b_2=c_2=0$.
The remaining linear equations are easily solved and yield
\begin{align}\label{Usquirmer}
U & =\zeta^{\perp}_h(a^2, \epsilon, \lambda)\, U_0\\
& = -\frac{3}{2N}(\lambda-1)\epsilon^3 a^2 + \zeta^{\perp}_h(0, \epsilon, \lambda). 
\end{align}
with
\begin{align}
 	N=&2\epsilon^5(\lambda-1) +3\lambda +2,\\  
N\zeta^{\perp}_h(0,\epsilon, \lambda) = &3\lambda +2 - (\lambda - 1)\epsilon^3 (3 \epsilon^2 -5)  
\end{align}


The offset of the squirmer from the center of the drop in a
direction perpendicular to its symmetry axis (see Fig.\ref{device})
gives rise to an angular velocity  $\bm{\omega}=\omega\bm{e}_y$ of the particle with
\begin{align}\label{eq:omegasquirmer}
\omega=-\kappa_h(\epsilon, \lambda)a U_0=-\frac{15\epsilon^3  (\lambda -1)}{2}aU_0 .
\end{align}
The angular velocity vanishes linearly with $a$.  \\

The drop moves in the direction of the symmetry axis of the
squirmer, $\bm{v}_{CM}=v_{CM}\bm{e}_z$, obtained from Eq.~\ref{eq:vcm}, with
\begin{align}\label{vCMsquirmer}
  v_{cm}= \mu^{\perp}_hU_0=\frac{5\epsilon^3\lambda}{N}U_0.
\end{align}

In I we considered an autonomous
swimmer, which is displaced from the center {\it parallel} to its symmetry axis $\mb{e}_z$. We now combine these results with our new ones for perpendicular
alignment to obtain the mobilities for general orientations of displacement $\bm{a}$ and symmetry axis of the squirmer $\bm{n}=\bm{U}_0/U_0$, and we write the  linear superposition of both results in coordinate free form as follows:
\begin{align}
\bm{U} & = \hat{\bm{\zeta}}_h \bm{U}_0  \label{eq:Ueqofmotion}\\
  \bm{v}_{cm} & = \hat{\bm{\mu}}_h \bm{U}_0 \label{eq:vcmeqofmotion}\\
  \bm{\omega}&=\kappa_h \bm{a}\times\bm{U}_0.\label{eq:omegaeqofmotion}  
\end{align}

The resulting mobility tensor $\hat{\bm{\zeta}}_h$ is symmetric and uniaxial with respect to the $\bm{a}$-direction
$\bm{e}_{||}=\bm{a}/a$, i.e.
\begin{equation}
\hat{\bm{\zeta}}_h= \zeta^{||}_h(a^2,\epsilon,\lambda) \bm{e}_{||}\otimes\bm{e}_{||} +
\zeta^{\perp}_h(a^2, \epsilon, \lambda)(\hat{\bm{1}}- \bm{e}_{||}\otimes\bm{e}_{||}).
\end{equation}
The longitudinal component  $\zeta^{||}_h$ 
follows from Eqs.(38,39) of I and is listed in
Appendix~\ref{app:mobilities}. The anisotropy vanishes trivially for $a=0$,
when the encapsulated particle is located at the center of the
drop~\cite{Lauga2017}.

The mobility of the drop turns out to be isotropic, i.e. $\mu^{\perp}_{h}=\mu^{||}_{h}=\mu_h$ and $\hat{\bm{\mu}}_h=\mu_h\bm{1}$.
Rotation of the drop is only observed for a displacement $\bm{a}$
with a nonzero component perpendicular to the symmetry axis $\bm{n}$
of the squirmer. The corresponding mobility tensor is uni-axial but anti-symmetric, so that it is determined by a single coefficient $\kappa_h$,
which can be read off from Eq.(\ref{eq:omegasquirmer}).

   \subsection{Passive particle dragged by an external force}

   Next, we consider a passive particle ($h=0$), which is dragged by
   an external force $\bm{F}_{ext}=F_{ext}\bm{e}_z$, perpendicular to
   its displacement $\bm{a}=a\bm{e}_x$ from the center of the
   drop. The coefficients $c_2,c_3$ are determined by the external
   force. Soving for the remaining coeffcients, we find
   \begin{align}
\bm{v}_{cm}=&v_{cm}\bm{e}_z=\mu_F^{\perp}\bm{F}_{ext}\\
     \bm{U}=&U\bm{e}_z=\zeta_F^{\perp}\bm{F}_{ext}\\
     \bm{\omega}=& \omega\bm{e}_y=\kappa_F\bm{a}\times\bm{F}_{ext}
   \end{align}
     The coeffcients $\mu_F^{\perp},\zeta_F^{\perp},\kappa_F$ are
     given in Eqs.~(\ref{eq:zetaFT}),(\ref{eq:muFT}),\ref{eq:kappaF}) of Appendix~\ref{app:mobilities}. A finite angular velocity of the particle
     is due to the torque (in frame D) exerted by the external force
     due to a finite displacement of the particle from the center of
     the drop.

     We proceed as in the previous subsection: we combine the above
     results for perpendicular alignment of $\bm{a}$ and
     $\bm{F}_{ext}$ with those of ref. I for parallel
     alignment. General orientations of $\bm{a}$ and $\bm{F}_{ext}$
     then give rise to mobility tensor relations, which read in
     coordinate free representation as follows:
\begin{align}
\bm{U} & = \hat{\bm{\zeta}}_F \bm{F}_{ext}  \label{eq:Udragged}\\
  \bm{v}_{cm} & = \hat{\bm{\mu}}_F \bm{F}_{ext} \label{eq:vcmdragged}\\
   \bm{\omega}&=\kappa_F \bm{a}\times\bm{F}_{ext}.\label{eq:omegadragged}  
\end{align}
The tensors $\hat{\bm{\zeta}}_F$ and $\hat{\bm{\mu}}_F$  are  symmetric and uniaxial with respect to the displacement $\bm{a}$. 
The longitudinal components follow from Eqs.(40,41) of I and are recalled in Appendix~\ref{app:mobilities}. 

 \subsection{ Passive particle subject to an external torque}

 Finally we consider a passive particle ($h=0$) with no applied force
 ($\bm{F}_{ext}=0$), subject to an external torque $\bm{D}_{ext}$ in frame P. To construct the most general case, we must discuss both  perpendicular and parallel alignment of
 torque and displacement, but the latter case has not yet been included in our discussion. It requires an extension of the calculations of reference I, which is given in Appendix {\ref{app:torque}}.

 For perpendicular alignment and in agreement with the coordinates chosen in
 sec.~\ref{analytics}, we choose $\bm{D}_{ext}=D_{ext}\bm{e}_y$, so that $\bm{\omega}=\omega\bm{e}_y$ and
 $\bm{U}=U\bm{e}_z$ . 
 In the absence of an applied
 external force, we have $b_3=c_3=0$. Torque balance determines the
 coefficient $c_2$ according to Eq.~(\ref{torquebalance}) and hence also
 $b_2=c_2/\lambda$ according to  Eq.~(\ref{bcIII.3}).  The transverse mobilities in the equations
 \begin{align}
U & = \zeta_D^\perp(a^2, \epsilon, \lambda) D_{ext}a\\
  v_{cm} & = \mu_D^\perp(a^2, \epsilon, \lambda) D_{ext}a\\
   \omega &=\kappa^\perp_D(a^2, \epsilon, \lambda) D_{ext}
 \end{align}
  are explicitly given in Eqs.(\ref{eq:zetaD},\ref{eq:muD},\ref{eq:kappaDT}) of Appendix{\ref{app:mobilities}}.

 The configuration with parallel alignment of torque and displacement leads to a spinning motion of the particle around its direction of propulsion. Its calculation 
 requires an extension of the analysis given in reference I, which is presented in Appendix \ref{app:torque}.  The result is $U=0$ and $\omega=\kappa_D^{||}D_{ext}$ with
 \begin{equation}\label{kappa_parallel}
\kappa_D^{||}= \frac{1}{8\pi\epsilon^3\eta^-}\big(
   \epsilon^3(\lambda-1) +1\big) 
 \end{equation}

 In coordinate free representation, the relations between $\bm{D}_{ext}$ and the velocities take on the form
\begin{align}
\bm{U} & = {\zeta}_D \bm{a}\times\bm{D}_{ext}  \label{eq:Utorque}\\
  \bm{v}_{cm} & = {\mu}_D \bm{a}\times\bm{D}_{ext} \label{eq:vcmtorque}\\
   \bm{\omega}&=\hat{\bm{\kappa}}_D \bm{D}_{ext},\label{eq:omegatorque}  
\end{align}
with a symmetric uniaxial tensor $\hat{\bm{\kappa}}_D$.

 \section{Discussion}
 \label{discuss}
The general mobility tensors are obtained by superposition of the
3 special cases worked out above and will now be used to discuss the general motion of drop and encapsulated particle.

\subsection{Motion of the drop}
 
A linear velocity of the drop is generated by all three driving
mechanisms: slip velocity of the squirmer, external force and external
torque
\begin{equation}
  \bm{v}_{cm}  =\mu_h\bm{U}_0+\hat{\bm{\mu}}_F \bm{F}_{ext} +\mu_D\bm{a}\times \bm{D}_{ext}
\end{equation}
The response to active slip, as characterized by $\mu_h$, is completely
isotropic and independent of $a$. In other words the linear velocity
of the drop with an encapsulated squirmer only depends on the size
$\epsilon$ of the squirmer and the viscosity contrast $\lambda$. For small $\epsilon$, it vanishes proportional to the volume of the squirmer.

If the drop is driven by an external force, acting on a passive encapsulated particle, then the response is anisotropic and characterized by the uniaxial tensor: $\hat{\bm{\mu}}_F= \mu^{||}_F(a^2) \bm{e}_a\otimes\bm{e}_a +
\mu^{\perp}_F(a^2)(\bm{1}- \bm{e}_a\otimes\bm{e}_a)$. The ratio of the two mobilities is given by
\begin{equation}
\frac{\mu^{||}_F}{\mu^{\perp}_F}=\frac{G-3a^2}{G-6a^2}\quad \text{with}\quad G=4\epsilon^4(\lambda-1)-5\epsilon^2+3(2\lambda+3)\nonumber
\end{equation}
so that the velocity of the drop is always larger for parallel
alignment of force and displacement.
The mobility of the drop remains finite as the size of the particle goes to zero and in fact coincides with the mobilities derived
previously~\cite{Rueckert2021b} for a point force inside a drop.
Finally, a torque exerted on the encapsulated particle, also propels
the drop, provided the particle is placed off center.

\begin{figure*}[htp]
\centering
\begin{subfigure}[]{}
\includegraphics[width=0.3\textwidth]{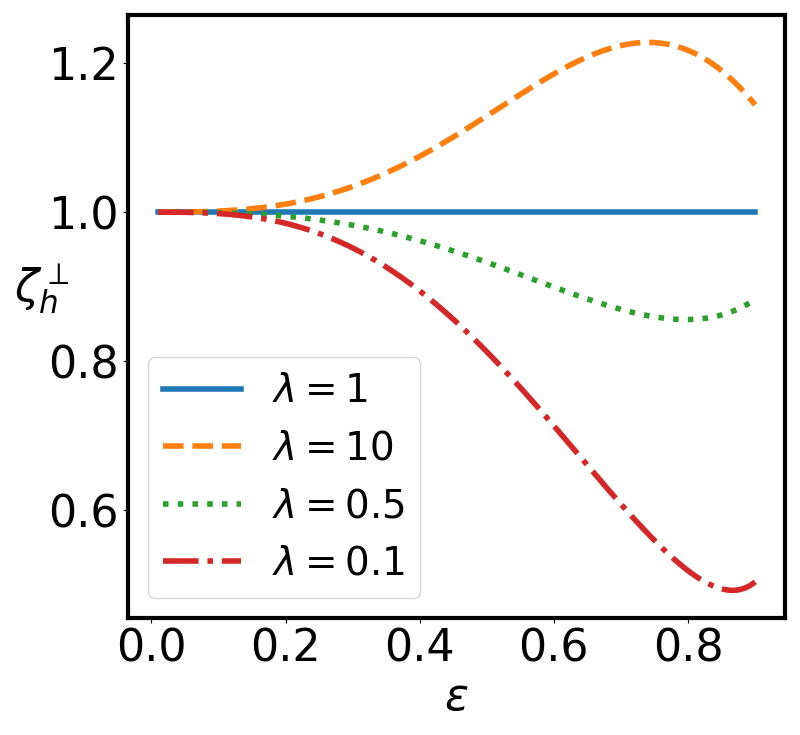}
\end{subfigure}
\begin{subfigure}[]{}
\includegraphics[width=0.3\textwidth]{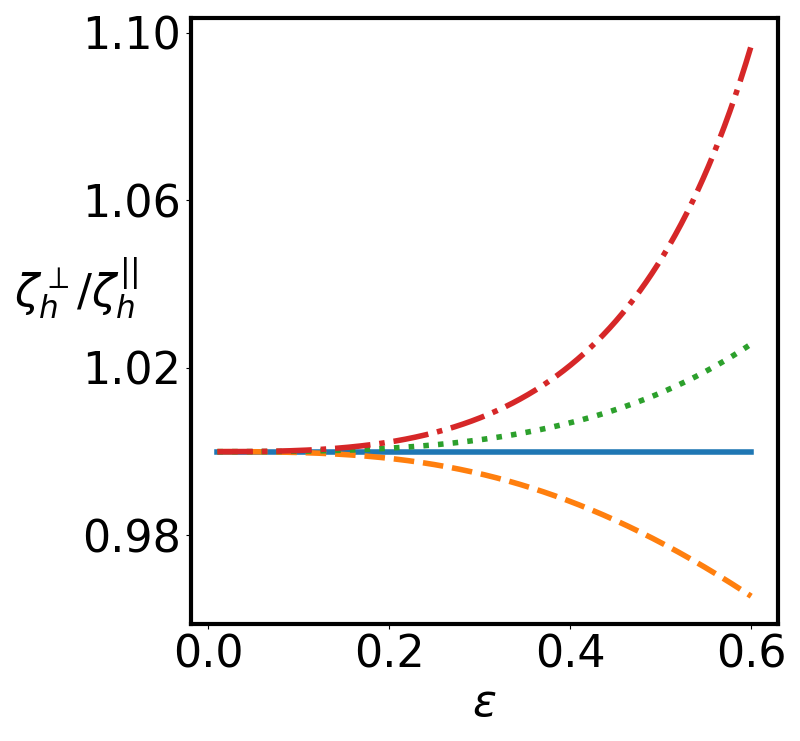}\end{subfigure}
\begin{subfigure}[]{}
\includegraphics[width=0.3\textwidth]{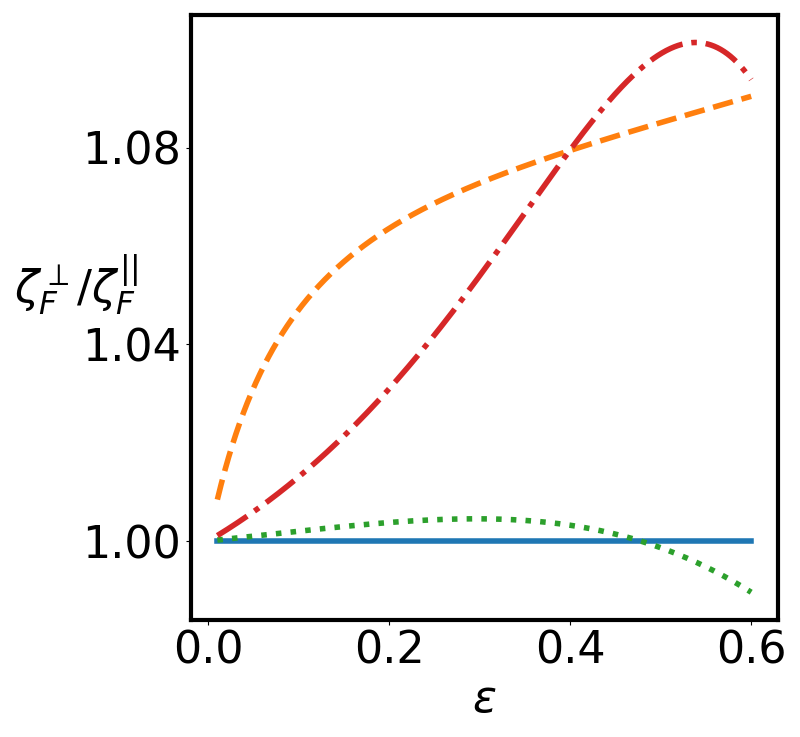}
\end{subfigure}
\caption{\label{fig:zetas} Mobility  of a squirmer (a) and mobility anisotropy of a squirmer (b), and a passive particle dragged by an external force (c), \textit{vs.} radius of the particle for different viscosity contrasts $\lambda$, as shown in the legend of (a). In (a) the distance of the particle from the drop's center is $r_s=0.1$, in (b) and (c) it is $r_s=0.4$.}
\end{figure*}

\subsection{Linear velocity of the particle}

The general propulsion velocity of the encapsulated particle is given by
\begin{equation}
\bm{U} =\hat{\bm{\zeta}}_h \bm{U}_0+\hat{\bm{\zeta}}_F \bm{F}_{ext} +{\zeta}_D \bm{a}\times\bm{D}_{ext}.
\label{eq:particlevelocity} 
\end{equation}
The response of the particle to either an applied force or to a
nonzero slip is in general anisotropic.
If only an active slip is present or only an external force is applied, the velocity is not in the
direction of the squirmer axis or the external force.   The reason for this anisotropy
is the response flow due to reflection at the interface, which is not
concentric around the particle. The anisotropy therefore vanishes as
$\epsilon\to 0$, or equivalently as the radius of the drop goes to
infinity. In that limit we recover the result for the squirmer in free
space, $\bm{U}= \bm{U}_0$, and the result for the mobility
of a passive particle dragged by an external force,
$\bm{U} =\frac{1}{6\pi\eta^{-}\epsilon} \bm{F}_{ext} $, in leading
order in $\epsilon$.

If  the squirmer fills
the whole drop ($\epsilon\to 1$), we also find $\bm{U}\to \bm{U}_0$, implying nonmonotonic
behaviour of $\hat{\bm{\zeta}}_h$ as a function of
$\epsilon$. In Fig.\ref{fig:zetas}a we show $\zeta_h^{\perp}$ as a function of $\epsilon$ for several values of $\lambda=\eta^-/\eta^+$ and $r_s=|\bm{r}_s|=0.1$.
One clearly observes nonmonotonic behaviour for all
$\lambda$. Furthermore, if the interior of the drop has a higher
viscosity than the outside ($\lambda>1$), the drop moves faster
than in free space, because the frictional forces in the interior are
larger than in the exterior region. The opposite behaviour is observed
for $\lambda<1$, i.e. a less viscous interior.  In
Fig.\ref{fig:zetas}b and c, we illustrate the anisotropy of the
response. The ratio $\zeta_h^{\perp}/\zeta_h^{||}$ is shown in
Fig.\ref{fig:zetas}b as a function of $\epsilon$ for several values
of $\lambda$ and $r_s=0.4$ in comparison to the isotropic case which
is realised for $\lambda=1$. Both cases,
$\zeta_h^{\perp}\lessgtr \zeta_h^{||}$ are possible, depending on
whether $\eta^+\lessgtr \eta^- $. 

If a passive particle dragged by $\bm{F}_{ext}$ fills the whole drop ($\epsilon\to 1$), one obtains $\bm{U} =\frac{1}{6\pi\eta^{+}\epsilon} \bm{F}_{ext}$, as expected. The dependence of the
anisotropy on the viscosity contrast $\lambda$ is more subtle, as can
be seen in Fig.\ref{fig:zetas}c, where we plot the ratio of
perpendicular to parallel mobility for an applied force,
$\zeta_F^{\perp}/\zeta_F^{||}$. For $\lambda <1$, the ratio becomes a non-monotonic function of $\epsilon$, and it may show both possibilities $\zeta_h^{\perp}\lessgtr \zeta_h^{||}$, as illustrated by $\lambda=0.5$ in Fig.{\ref{fig:zetas}c.


\subsection{Rotational velocity of the particle}

All 3 driving mechanisms, slip, external force and external torque,
give rise to a rotation of the particle:
\begin{equation}
  \bm{\omega} =\kappa_h\bm{a}\times \bm{U}_0+\kappa_F\bm{a}\times \bm{F}_{ext}+
  \hat{\bm{\kappa}}_D \bm{D}_{ext}, 
  \label{eq:particlerotation}  
\end{equation}
The rotational mobility due to slip vanishes as the volume of the
squirmer, i.e. there is no rotational motion of a squirmer in free
space. Note that we have not included a chiral component of the slip
which has been discussed for a squirmer in unbounded
space~\cite{Lauga2014}.

An external torque causes a rotation of the particle with in general anisotropic mobilities $\kappa_D^{||}\neq\kappa_D^{\perp}$. In free space, i.e. in the limit $R\to \infty$, the response becomes isotropic and reduces to the rotational mobility in free space: $\kappa_D^{||}=\kappa_D^{\perp}=\frac{1}{8\pi\eta^-\epsilon^3}$.


\section{Conclusions and outlook}\label{conclusions}

We have analysed the dynamics of a solid particle, encapsulated in a
drop and displaced from the drop's center by a general vector
$\bm{a}$ (non-axisymmetric configuration). Several
driving mechanisms have been considered. Either the solid particle is
a (uniaxial) squirmer, driven by an active slip or it is subject to an
external force or to an external torque, or any combination thereof.
We have derived analytical expressions for the translational and
rotational mobilities, i.e. the linear and rotational velocity of the
squirmer as well as the linear velocity of the drop as functions of
translation vector $\bm{a}$, particle radius $\epsilon$ and viscosity
contrast $\lambda=\eta^-/\eta^+$. 
Our analytical method is adapted to mobility problems in spherical geometries, for which it  is simple and straightforward. It can easily be generalized to  more complex squirmers, which possess chiral components and/or higher $l$-components of active slip velocity. 
The obtained results provide a first step towards controlled locomotion of an
(active) particle, encapsulated in a spherical liquid drop. Based on the
general results for the linear ($\bm{U}$) and rotational ($\bm{\omega}$) velocity of the particle as well as the linear velocity of the drop ($\bm{v}_{CM}$),
one has to solve the equations of motion for the particle in the rest
frame of the drop:
\begin{align}
  {\dot{\bm{r}}_s}&=\bm{u}=\bm{U}-\bm{v}_{CM}\label{rsdot}\\
  {\dot{\bm{n}}}&=\bm{n}\times\bm{\omega}\label{ndot}.
  \end{align}
  Together with the equation for
  $\bm{v}_{CM}$, one thereby obtains the trajectories of drop and
  squirmer. Adjusting the external force and torque, should allow to
  steer the composite system to designed places, as required by drug
  delivery or more generally in the context of microrobotics.

\appendix

\section{Mobility tensors}
\label{app:mobilities}
We now give the explicit form of all the mobility tensors of the encapsulated particle and the drop as functions of $a, \eta^\pm$ and $\epsilon$.  There are 5 symmetric tensors, 
	$\bm{\zeta}_w, \bm{\mu}_w, \bm{\kappa}_D$ (with $ \, w=h,F$), and 4 anti-symmetric tensors characterized by $\kappa_w, \zeta_D, \mu_D$. All these functions are polynomials of $a^2$ of degrees up to 2. The parallel components of $\bm{\zeta}_w$ and $\bm{\mu}_w$ have been obtained in I. They are included here for completeness. All other components are calculated by solving the linear system of equations set up in sect. (\ref{bcsquirmer}) and  sect.(\ref{bcdrop})  by symbolic computing using \textit{SymPy} \cite{sympy} (except the mobility $\kappa^{||}_D$, which is obtained in Appendix \ref{app:torque}).
\\

1.) $\bm{U}=\bm{\zeta_h}\bm{U}_0$
\begin{align}
	N\zeta^{||}_h &= 3\epsilon^3(\lambda-1)a^2 + C_{h0}\label{eq:zetahP}\\
	N\zeta^\perp_h &= -\frac{3}{2}\epsilon^3(\lambda-1)a^2 +C_{h0}\label{eq:zetahT}
\end{align}
with
\begin{equation}
	C_{h0}=3\lambda +2 -\epsilon^3(\lambda-1)(3\epsilon^2-5)
\end{equation}
\\
2.) $\bm{U}=\bm{\zeta_F}\bm{F}_{ext}$
\begin{align}
	N\zeta^{||}_F &=\frac{1}{6\pi\epsilon\eta^-}\big(C^{||}_4a^4  +C_{F0}\big)\label{eq:zetaFP}\\	
	N\zeta^{\perp}_F &=\frac{1}{6\pi\epsilon\eta^-}\big(C^{\perp}_4a^4  + C^{\perp}_2a^2 +C_{F0}\big)\label{eq:zetaFT}
\end{align}
with
\begin{align}
	C_{F0}= &  \epsilon(\lambda-1)\Big(2\epsilon^5(\lambda-1)+\frac{9}{2}\epsilon^4-5\epsilon^2\\\nonumber
	& +\frac{3}{2}(2\lambda + 3) \Big) +3\lambda +2
\end{align}
\begin{align}
	C^{||}_4 = & -\frac{9\epsilon}{10}(\lambda-1)\\
	\end{align}
and
\begin{align}
	C^{\perp}_4 = &-C^{||}_4\\
	C^{\perp}_2= &  \frac{3(\lambda-1)}{4}\Big(2\epsilon^6(\lambda-1)+5\epsilon^3\\\nonumber
	& +3\epsilon(\lambda-1)\Big)\\
\end{align}
\\
3.)  $\bm{U}=\zeta_D \bm{a}\times\bm{D}_{ext}$
\begin{equation}
	N\zeta_D=\frac{\lambda-1}{8\pi\eta^-}\left(\frac{3}{2}a^2 + C_D\right)\label{eq:zetaD}
\end{equation}
with 
\begin{equation}
	C_D=2\epsilon^5(\lambda-1) + 5\epsilon^2 + 3(\lambda-1)
\end{equation}
\\
4.) $\bm{v}_{cm}=\bm{\mu}_h\bm{U}_0$

\begin{align}
	N\mu^{||}_h = N\mu^{\perp}_h=&5\lambda\epsilon^3\label{eq:muh}
\end{align}
\\
5.)  $\bm{v}_{cm}=\bm{\mu}_F\bm{F}_{ext}$

\begin{align}
	N\mu^{||}_F= &\frac{1}{12\pi\eta^+}\Big(-3a^2 +G\Big)\label{eq:muFP}\\
	N\mu^{\perp}_F= &\frac{1}{12\pi\eta^+}\Big(-6a^2 +G\Big)\label{eq:muFT}
\end{align}
\begin{align}
G = & 4\epsilon^5(\lambda-1) - 5\epsilon^2 + 3(2\lambda+3)	
\end{align}
\\
6.) $\bm{v}_{cm}=\mu_D \bm{a}\times\bm{D}_{ext}$

\begin{align}
	 N\mu_D= \frac{-5}{8\pi\eta^+}\label{eq:muD}
\end{align}
\\
7) $\bm{\omega}= \kappa_h\bm{a}\times\bm{U}_0$
\begin{align}
	N\kappa_h=\frac{15}{2}\epsilon^3(\lambda-1)\label{eq:kappah}
\end{align}
\\
8) $\bm{\omega}= \kappa_F\bm{a}\times\bm{F}_{ext}$
\begin{align}
	N\kappa_F= -\frac{\lambda-1}{8\pi\eta^-}(6a^2+K)\label{eq:kappaF}
\end{align}
with
\begin{equation}
	K=2\epsilon^5(\lambda-1) +5 \epsilon^2 +3(\lambda-1)
\end{equation}
\\
9)  $\bm{\omega}= \bm{\kappa_D}\bm{D}_{ext}$
\begin{align}
	\kappa^{||}_D= &\frac{1}{8\pi\eta^-}M\label{eq:kappaDP}\\
	\kappa^{\perp}_D= & \frac{1}{8\pi\eta^-}\Big(\frac{15a^2}{2N}(\lambda-1)+M \Big)\label{eq:kappaDT}
\end{align}
with
\begin{align}
	M= (\lambda-1) + \epsilon^{-3}
\end{align}

\section{Parallel alignment of displacement and torque \label{app:torque}}

In I, we analysed two  uniaxial configurations: a
squirmer displaced from the center of the drop such that its
symmetry axis coincides with the direction of the displacement and a
passive particle displaced from the center of the drop such that
the applied force is parallel to the displacement. Here we discuss the
extension to an applied torque, parallel to the displacement. To make
use of the formalism developed in I, we choose
$\bm{D}_{ext}=D\bm{e}_z$ and consider a displacement
$\bm{a}=a\bm{e}_z$. We expect a rotational velocity of the particle
$\bm{\omega}=\omega\bm{e}_z$ and possibly a linear velocity
$\bm{U}=U\bm{e}_z$. The flow field on the surface of the particle is
given by
\begin{align}
  \bm{v}^-(\bm{r})=&\bm{U}+\bm{\omega}\times \bm{r}\nonumber\\
                  =& U\bm{e}_z+r\omega\sin{\theta}\bm{e}_{\phi}\label{eq:bc1torque}  
  \quad \text{for} \quad \bm{r}\in \partial V_s. 
\end{align}
The contribution, due to the rotation, can be expressed in terms of
vector spherical harmonics,
$r\omega\sin{\theta}\bm{e}_{\phi}=-r\omega\bm{\Psi^2_{10}} $, and
gives rise to a corresponding component of the flowfield, $\bm{u}^1_{02}$. We thus have to extend
Eqs.(10-13) of I

\begin{eqnarray}
  \label{l=1flowtorque}
\bm{u}^{1<}&:=& \bm{u}^{1<}_{10}= \bm{\Psi}^1\\
\bm{u}^{2<}&:=&\bm{u}_{10}^{2<}=\frac{r}{2}\bm{\Psi}_{10}^2\\
  \bm{u}^{3<}&:=& \bm{u}^{3<}_{10}=                  \frac{r^2}{3}
                 (\bm{\Psi}^1+\frac{1}{5} \bm{\Psi}^3)\\
  \bm{u}^{1>}&:=& \bm{u}^{1>}_{10}=\frac{1}{r^3}  \bm{\Psi}^3\\
\bm{u}^{2>}&=&\bm{u}_{10}^{2>}=-\frac{1}{r^{2}}  \bm{\Psi}^2_{10}\\
  \bm{u}^{3>}&:=&\bm{u}^{3>}_{10}=\frac{1}{3r} (-\bm{\Psi}^3
                      +4\bm{\Psi}^1) . 
\end{eqnarray}
The general solution of Stokes equation inside the drop is thus given by
\begin{equation}
  \label{eq:v-app}
  \bm{v}^-=a_1\bm{u}^{1<}+a_2 \bm{u}^{2<}+a_3\bm{u}^{3<}+b_1\bm{u}^{1>}
  +b_2\bm{u}^{2>}+b_3\bm{u}^{3>},
\end{equation}
and the flow field outside of the drop by
\begin{equation}
\label{eq:exteriorvelocity_app}
  \bm{v}^+=c_1\bm{u}^{1>}++c_2 \bm{u}^{2>}+ c_3\bm{u}^{3>}.
\end{equation}
Translations in the z-direction have been worked out in I for $\bm{u}^{1<},\bm{u}^{3<},\bm{u}^{1>},\bm{u}^{3>}$ and are easily extended to the 2 new components:
\begin{eqnarray}
  \label{l=1m=0}
   \bm{u}^{2<}(\bm{r}-a\bm{e}_z)&=&\bm{u}^{2<}(\bm{r})\\
  \bm{u}^{2>}(\bm{r}-a\bm{e}_z)&=&\bm{u}^{2>}(\bm{r})+{\cal O}(l\geq 2)
 \end{eqnarray}
The complete solution is now substituted into the boundary condition to determine the yet unknwon coeffcients. On the surface of the squirmer, Eq.~(\ref{eq:bc1torque})  implies
\begin{align}
  a_1+\frac{\epsilon^2}{3}a_3+\frac{4}{3\epsilon}b_3=& \,U \label{coeff1}\\
   \frac{\epsilon^2}{15}a_3+\frac{1}{\epsilon^3}b_1-\frac{1}{3\epsilon}b_3=&\, 0.\label{coeff2}\\
    \frac{\epsilon}{2}a_2-\frac{b_2}{\epsilon^2}=&-\epsilon\omega \label{coeff3}              \end{align}
On the surface of the drop we require continuity of the flow
\begin{align}
  a_1+\frac{a_3}{3}(1+\frac{3}{5} a^2)+\frac{4}{3}(b_3-c_3)&=0 \label{coeff4}\\
 b_1-\frac{b_3}{3}(1-\frac{3}{5} a^2)+\frac{a_3}{15}-c_1+\frac{c_3}{3}&=0\label{coeff5}\\
   \frac{a_2}{2}-b_2+c_2&=0 . \label{coeff6}                               
\end{align}

and continuity of  the tractions
  \begin{align}
  \frac{a_3}{10}-b_1-\frac{a^2}{5} b_3+\frac{c_1}{\lambda}&=0\label{coeff7}\\
    c_3-\lambda b_3&=0\label{coeff8} \\
  3\eta^-  b_2&= 3\eta^+ c_2.\label{coeff9}
  \end{align}

  These are 9 equations, which together with force and torque balance
  determine the 9 coefficients in the general ansatz for the flow and
  $U$ and $\omega$. The equtions simplify considerably in this case:
  $c_3=b_3=0$ (force balance), $c_2=D/(8\pi\eta^+)$ is determined by the external
  torque and $\lambda b_2=c_2$ (Eq.\ref{coeff9}).  This leaves us with
  5 equations (\ref{coeff2}), (\ref{coeff4}), (\ref{coeff5}),(\ref{coeff6}), (\ref{coeff7}) for the
  remaining 5 coefficients ($a_1,a_2,a_3,b_1,c_1$), one equation for
  $U$ (Eq.\ref{coeff1})and one equation for $\omega$
  (Eq.\ref{coeff3}). The result is $U=0$, i.e. no translational velocity of the squirmer; furthermore $a_1=a_3=b_1=c_1=0$ and
  \begin{equation}
    \omega=\frac{D}{8\pi\epsilon^3\eta^-}\big(
   \epsilon^3(\lambda-1) +1\big) = \kappa^{||}_D D.
\end{equation}
This result is used in Eq.(\ref{kappa_parallel}).


  \bibliographystyle{iopart-num} \bibliography{squirmer}

 \end{document}